\documentclass{article}
\usepackage{ijcai22}

\usepackage{times}

\usepackage{soul}
\usepackage{url}
\usepackage[hidelinks]{hyperref}
\usepackage[utf8]{inputenc}
\usepackage[small]{caption}
\usepackage[pdftex]{graphicx}
\usepackage{amsthm,amsmath,amssymb}
\usepackage{multirow}
\usepackage{diagbox}
\usepackage{booktabs}
\soulregister{\cite}7

\usepackage{color, xcolor}
\usepackage{xspace}

\title{Learnable Model Augmentation Self-Supervised Learning \\
for Sequential Recommendation}

\author{
 Yongjing Hao$^1$\and
 Pengpeng Zhao$^1$\footnote{Contact Author}\and
 Xuefeng Xian$^2$\and
 Guanfeng Liu$^{3}$\and
 Deqing Wang$^4$\and\\
 Lei Zhao$^1$\and
 Yanchi Liu$^5$\And
 Victor S. Sheng$^6$
 \\
 \affiliations
 $^1$School of Computer Science and Technology, Soochow University, Suzhou, China\\
 $^2$School of Computer Engineering, Suzhou Vocational University, Suzhou, China\\
 $^3$School of Computing, Macquarie University, Sydney, Australia\\
 $^4$School of Computer, Beihang University, Beijing, China\\
 $^5$Rutgers University, New Jersey, USA\\
 $^6$The University of Central Arkansas, Conway, USA
 \emails
 yjhaozb@stu.suda.edu.cn,
 ppzhao@suda.edu.cn,\\
 xianxuefeng@jssvc.edu.cn,
 guanfeng.liu@mq.edu.au,\\
 dqwang@buaa.edu.cn,
 zhaol@suda.edu.cn,\\
 yanchi@nec-labs.com,
 victor.sheng@ttu.edu
}

\begin{document}
\maketitle


\begin{abstract}
Sequential Recommendation aims to predict the next item based on user behaviour. Recently, Self-Supervised Learning (SSL) has been proposed to improve recommendation performance. However, most of existing SSL methods use a uniform data augmentation scheme, which loses the sequence correlation of an original sequence. To this end, in this paper, we propose a \textbf{L}earnable \textbf{M}odel \textbf{A}ugmentation self-supervised learning for sequential \textbf{Rec}ommendation \textbf{(LMA4Rec)}. Specifically, LMA4Rec first takes model augmentation as a supplementary method for data augmentation to generate views. Then, LMA4Rec uses learnable Bernoulli dropout to implement model augmentation learnable operations. Next, self-supervised learning is used between the contrastive views to extract self-supervised signals from an original sequence. Finally, experiments on three public datasets show that the LMA4Rec method effectively improves sequential recommendation performance compared with baseline methods.
\end{abstract}

\section{Introduction}
Sequential Recommendation (SR) has been widely employed in our daily life, such as Amazon, Tmall and Yelp platforms. Users' intents often change over time, making it difficult for these platforms to make appropriate recommendations. Various methods have been proposed by modeling users' dynamic interests for sequential recommendation. 

Early works using Markov Chain (MC) \cite{he2016fusing} to model the pair-wise item transition correlations for recommendation. Then, Factorization Personalized Markov Chain (FPMC) \cite{rendle2010factorizing} directly uses item transition matrices for recommendation. Next, the Fusing similarity models with Markov chains (Fossil) \cite{Fossil} uses similar items to deal with data sparsity issues in recommendation. With the development of deep neural networks, such as Recurrent Neural Networks (RNNs)\cite{zhao2020go}, which uses the last hidden state as user embeddings to predict the next interaction item about a user. The variants of RNN, such as Gated Recurrent Unit (GRU) \cite{hidasi2015session} and Long Short-Term Memory (LSTM) \cite{zhu2017next}, explore both short and long-term interaction item transition. Later, several methods were proposed with self-attention networks. For example, Self-Attentive Sequential Recommendation (SASRec) \cite{kang2018self} uses a transformer layer to learn item importance in sequences and achieves an excellent recommendation effect. 
However, supervised signals used in supervised recommendation methods can be obtained from observed interactions. Such interactions are very sparse compared to the entire interaction space of recommendation, leading to secondary recommendation results.

Recently, Self-Supervised Learning (SSL) obtains exciting performance, which aims to dig out information from unlabeled data. SSL is designed to minimize the consistency among different views of the same user or item. For instance, both BPR loss \cite{rendle2012bpr}, and NCE loss \cite{liu2021social}, which are widely used in recommended systems, use pairwise compare loss for positive and negative samples. There have been some SSL-based works in recommendation systems. For example, $S^3$-Rec \cite{zhou2020s3} uses attributes and random masks of items to maximize the mutual information between items, whereas SSL is used in its item encoder. CLS4Rec \cite{xie2020contrastive} is a SSL model to improve sequential recommendation, which uses random data augmentation operations. More recently, CoSeRec \cite{liu2021contrastive} uses item correlation to generate high-confidence positive sample pairs and achieves the most advanced recommendation result. However, those methods lose the sequence correlation of the original sequence.

To this end, we propose a new model called Learnable Model Augmentation Self-Supervised Learning for sequential recommendation (LMA4Rec). We study model augmentation as a supplementary data augmentation method. LMA4Rec has following advantages: \textbf{1)} The process of data augmentation of original data can be regarded as the process of injecting a data center into the disturbance. This operation can improve its robustness. \textbf{2)} Model augmentation does not perform any operators on the original data. It can retain the pattern and semantic information of the original sequence so that generated positive sample pairs have a higher degree of confidence. \textbf{3)} Joint learning makes LMA4Rec an end-to-end method that can generate high-quality contrastive views without manual operation. LMA4Rec contains three key components: a learnable model augmentation component, a self-supervised learning component, and a joint learning component. In short, the learnable model augmentation component first generates two high-quality comparison views of the original data. Second, the self-supervised learning component uses the contrast loss training model. Finally, the joint learning component effectively optimized the sequential recommendation and the self-supervised learning tasks. Experiments have verified the efficacy of LMA4Rec. 

The main contributions are summarized as follows:

\begin{itemize}
    \item To the best of our knowledge, It is the first effort that introduces learnable model augmentation to generate two self-supervised views.
    \item We propose a Learnable Model Augmentation Self-Supervised Learning (LMA4Rec), which integrates learnable model augmentation and self-supervised learning to improve performance. 
    \item  We conduct various experiments on real-world public datasets. Our experimental results demonstrate that LMA4Rec outperforms baselines.
\end{itemize}

\section{Related Work}
In this section, we briefly review related work, including sequential recommendation and self-supervised learning.
\subsection{Sequential Recommendation}
Previously, sequential recommendation usually models the sequential patterns with MC \cite{he2016fusing}. With the recent advance of the deep neural network, like Recurrent Neural Networks (RNNs) \cite{wu2017recurrent,zhao2020go} and its variants are applied for user behaviour sequences, such as LSTM \cite{he2016fusing}, and GRU \cite{hidasi2015session}. These methods can treat the processed data sequence as a time sequence. In addition to RNNs, methods based on Convolutional Neural Networks (CNNs) have also been proposed, like Convolutional Sequence Embedding Recommendation Model (Caser) \cite{tang2018personalized}, in which the sequence patterns about neighbours of the relevant user are captured with horizontal and vertical filters. Recently, Self-Attention has demonstrated outstanding ability in modelling sequence data. \cite{kang2018self} used SA to gain the different weights about the various items adaptively. On the other hand, Graph Neural Networks (GNNs) has aroused great attention. For instance, to capture both local and global structures, the Graph Contextualized Self-Attention Network (GC-SAN) \cite{xu2019graph} method combines self-attention with GNN and achieves promising results. 
However, the supervised signal in the supervised recommendation methods comes from the observed interaction, which is very sparse compared to the entire interaction space and leads to secondary recommendation results.

\subsection{Self-Supervised Learning}
Self-Supervised Learning (SSL) achieved satisfactory results in different fields. Such as Natural Language Processing (NLP) \cite{gao2021simcse,yang2019xlnet}, Computer Vision (CV) \cite{chen2020simple,he2020momentum}, Graph Embedding (GE) \cite{you2020graph,wu2021self} and Recommendation Systems (RS) \cite{xie2020contrastive}. Existing work has developed different SSL schemes from generation or contrast. For generative SSL, the masked language model is used in BERT \cite{devlin2018bert} to generate the masked words in the sentence. Gated pixel CNN \cite{oord2016conditional} generates a large number of pictures by modeling the probability distribution of pixels in CV. Compared with the generative SSL, the contrastive SSL scheme performs better. SimCLR \cite{chen2020simple} proposes a simple contrast learning between image augmentation views, which is quite effective in implementing SSL. GraphCL \cite{you2020graph} uses comparative learning between views from corrupted graph structures. CL4SRec \cite{xie2020contrastive} and CoSeRec \cite{liu2021contrastive} designed SSL sequence augmentation methods based on sequence recommendations. 
However, these existing self-supervised learning methods are data-based augmentation. We argue that data-based augmentations suffer from losing the pattern and semantics of the original sequence.

\begin{figure*}[t]
	\centering
	\includegraphics[width=0.75\textwidth]{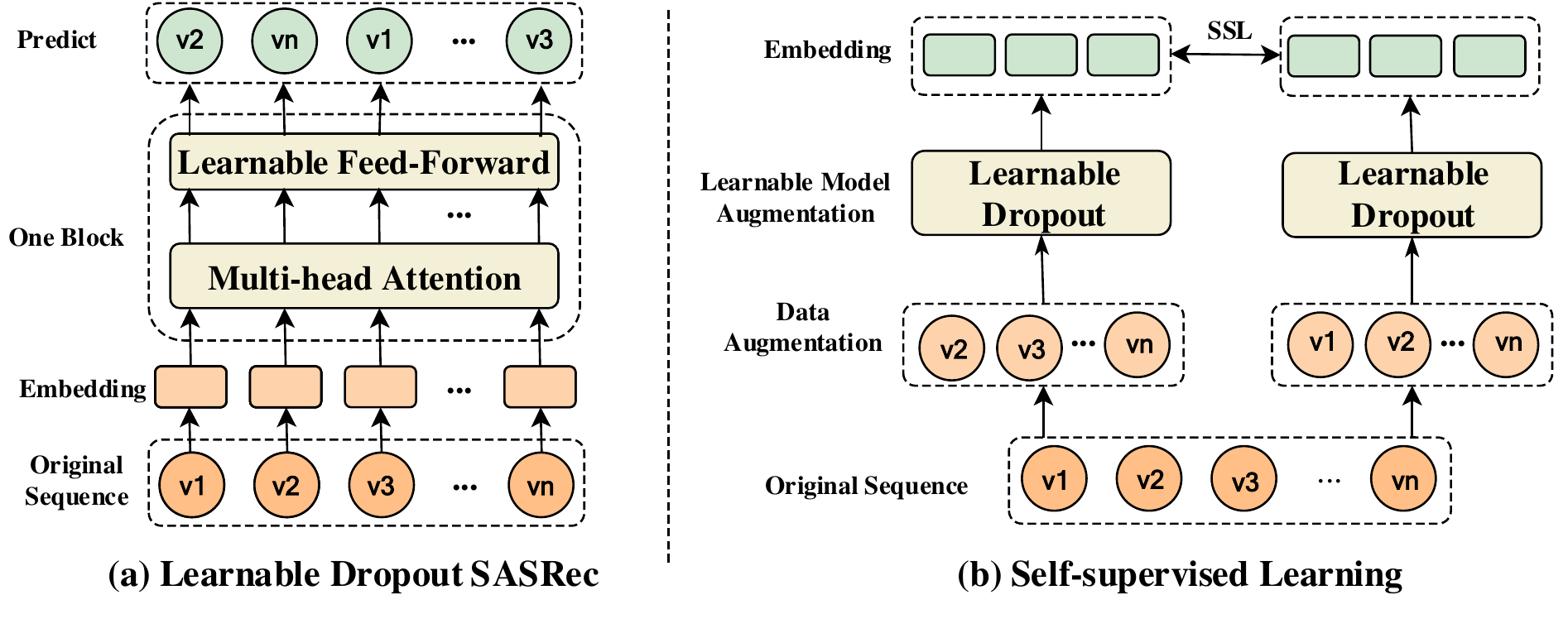}
	\vspace{-0.1cm}
	\caption{
	(a) The model of learnable dropout SASRec.
    (b) The model of self-supervised contrastive learning with learnable model augmentation, which can constructs two views for self-supervised learning.
	 } \label{fig1}
\end{figure*}

\section{Preliminaries}

\textbf{Problem Formulation}
Before explaining our LMA4Rec, we first give the definition of the sequential recommendation problem. The representation of all user sets is $U =\left \{ u_1, u_2,..., u_M \right \} $, the representation of all items sets is $I =\left \{ i_1, i_2,..., i_N \right \} $, where the numbers of all users and left and right items are expressed as $M$ and $N$. Next, we sort the items visited by a single user according to time. This sequence is expressed as $ S =\left \{s_1, s_2,..., s_{|S|} \right \} $. Sequential recommendation aims to predict the item that the user may interact with at the next moment based on the user's historical interaction sequence. 

\textbf{Theorem of Augment-Reinforce-Merge (ARM).}
For a vector of $N$ binary random variables $Z=( Z_1,Z_2,...,z_N)^T$, the gradient of $Z$ is \cite{yin2018arm}:
\begin{equation}
\label{eq.1}
\varepsilon \left ( \phi  \right ) = \mathbb{E}_{z\sim  {\textstyle \prod_{n=1}^{N}}Bernoulli (z_n;\sigma(\phi _n))}[f(z)],
\end{equation}
with respect to $\phi =(\phi_1,\phi_2,...,\phi_N)^T$, the logits of the Bernoulli probability parameters can be expressed as:
\begin{equation}
\begin{aligned}
\label{eq.2}
\bigtriangledown_\phi \varepsilon (\phi )=\mathbb{E}_{u\sim {\textstyle \prod_{n=1}^{N}}Uniform(u_n;0,1)}\\
[(f(1_{[u> \sigma(-\phi )]})-f(1_{[u< \sigma(\phi)]))(u-{1}/{2})}],
\end{aligned}
\end{equation}
where $1_{[u> \sigma(-\phi)]}: = {(1_{[u_1> \sigma(-\phi_1)]},...,1_{[u_N> \sigma(-\phi_N)]})}^T$, and $\sigma (\cdot )$ is the sigmoid function.

\section{The proposed method}
This section explains our LMA4Rec model for sequential recommendation in detail, which only uses the user's historical interact behaviour information. We first introduce a sequential recommendation model named Learnable Dropout SASRec. Secondly, we present a general self-supervised learning model and the specific operations of learnable model augmentation operations. Finally, we propose training the model through a joint learning model.

\begin{figure}[ht]
	\centering
	\includegraphics[width=0.45\textwidth]{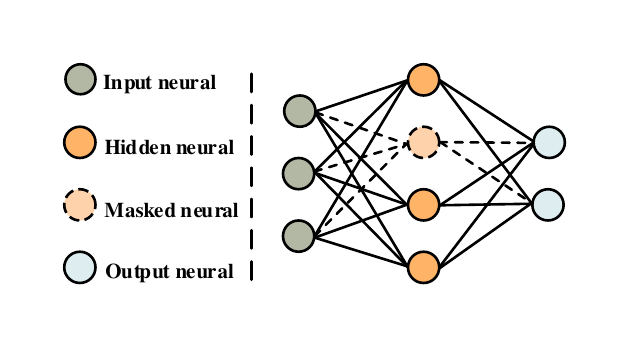}
	\vspace{-0.6cm}
	\caption{ Learnable Dropout.} \label{fig2}
\end{figure}

\subsection{Learnable Dropout SASRec}
Self-Attentive Sequential Recommendation (SASRec) is a powerful method to encode sequences, leading to successful application in sequential recommendation. SASRec includes two key components: a multi-head self-attention layer and a position-wise Feed-Forward Network (FFN) layer. The former aims to model project dependencies in order. The latter aims to embed the corresponding next item in the prediction sequence for each position. The sequential recommendation model in our work is improved based on SASRec. The embedding layer and self-attention block layers are the same as SASRec. Because DNNs often with over-fitting problems, to deal with this problem, Dropout as a standard regularization technology has a wide range of applications in DNNs. The existing Dropout randomly masks neurons on the training set with a certain probability and adopts all neurons on the test set. We believe that using the same dropout rate for different layers may lead to secondary results. Inspired by \cite{boluki2020learnable}, we introduce the learnable Bernoulli dropout into SASRec. As for stacking self-attention blocks, we use the following operations to deal with the problem:
\begin{equation}
\label{eq.3}
d'(a) =x+\mathrm {LBD}(d(LayerNorm(a))),
\end{equation}
where $d(a)$ is the FFN. For the feed forward network layer in each block, we normalize the input $a$ before sending it to $d$, and apply Learnable Bernoulli Dropout (LBD) on the output of $d$, and then add the input $a$ to the final output.

\textbf{Learnable Bernoulli Dropout (LBD):} A sequense was given $S={(x_i, y_i)}_{i=1}^{Q}$, where $x$ denote the input item and $y$ denote the target item of users, a function $f(x;\alpha)$ from the input space to the target space with parameters $\alpha$. $f(x;\alpha)$ is a neural network function such as SASRec with the parameters $\alpha$. To learn the value of parameters $\alpha$, we can minimize the sequential recommendation loss function $L_{rs}$. In this paper, we use Log-likelihood Loss function without any regularization terms $R$. Using Stochastic Gradient Descent (SGD):
\begin{equation}
\label{eq.4}
L(\alpha |S)\approx \frac{N}{Q} \sum_{i=1}^{Q} \varepsilon (f(x_i;\alpha ) ),
\end{equation}
where $Q$ is the size of mini-batch. 

Assuming SASRec model with $C$ layers of FFN. As for ${(c- 1)}^{th}$ FNN layer the input is $K_{c- 1}$ neurons, as for $c^{th}$ FFN layer the output is $K_c$ neurons. We use the weight matrix $W_c \in R^{K_{c- 1} \times K_c}$  represent the connection $c- 1$ FFN layer to $c$ FFN layer. We add the dropout to the output for every FFN layer, which multiplies it with a random variable $z_c \sim p(z_c)$ element-wise. The most common choice for $p(z_c)$ is the Bernoulli distribution $Bern(\eta (\beta  c))$ with dropout rate $1- \eta (\beta  c)$, where we have reparameterized the dropout rate using the sigmoid function $\eta ( \cdot )$. Here the set of logits of all dropout parameters is represented by $\beta  = {\left \{ \beta c \right \}}_{c=1}^{C}$, and the set of all dropout masks is represented by $\beta  = {\left \{ Z_c \right \}}_{c=1}^{C}$. The dropout, as mentioned above, the processing method is one of the most commonly used regularization techniques in deep neural networks. This regularization technique can improve the algorithm's robustness to a certain extent while avoiding the problem of over-fitting. The dropout operation can finally be regarded as the weight of the data set itself. If the input data $z_{c_k} = 0$, the value of the weight matrix $W_c$ $k_{th}$ is also $0$. In the research of the existing literature, the parameters $\beta $ in the random mask matrix $z$ are treated as hyper-parameters, and the existing solution methods are implemented by grid search. Unlike other methods, in this article, the expectations of the dropout optimization loss function for Bernoulli distribution are as follows: 
\begin{equation}
\begin{aligned}
\label{eq.5}
\alpha =_{\left \{ \alpha \setminus \beta ,\beta  \right \} }^{min} \mathbb{E} \sim  {\textstyle \prod_{i=1}^{M}}Bern\left ( z_i;\eta \left ( \beta  \right )  \right )  \left [ L\left ( \alpha ,z|S \right )  \right ].
\end{aligned}
\end{equation}

Figure \ref{fig1}(a) shows the learnable dropout SASRec model. We denote a learnable dropout SASRec encoder as $SeqEnc( \cdot )$, an arbitrary sequence encoder. We formulate the encoding process as follows:
\begin{equation}
\label{eq.6}
h_u=SeqEnc(s_u)
\end{equation}
where $h_u$ denotes the sequence embedding of $s_u$. $h_u$ is a bag of embeddings in Transformer. For each position $t$, $h^t_u$ represents a predicted next-item. We adopt the log-likelihood loss function to optimize the encoder for next-item prediction as follows:
\begin{equation}
\begin{aligned}
\label{eq.7}
L_{rs}=-log\frac{exp(s_{u,t}^{\top }v_{t+1}^{+})}{exp(s_{u,t}^{\top }v_{t+1}^{+})+ {\textstyle \sum_{v_{t+1}^{-}}exp(s_{u,t}^{\top }v_{t+1}^{-})}},
\end{aligned}
\end{equation}
where $L_{rs}$ denotes the loss score for the prediction at position $t$ in sequence $s_u$.

\subsection{Self-Supervised Learning model}
The Self-supervised learning model includes four major layers, i.e., data augmentation layer, learnable model augmentation layer, user embedding layer and self-supervised loss layer. Its model is illustrated in Figure \ref{fig1}(b). The detailed explanation of each layer is as follows.

\begin{figure}[t]
	\centering
	\includegraphics[width=0.45\textwidth]{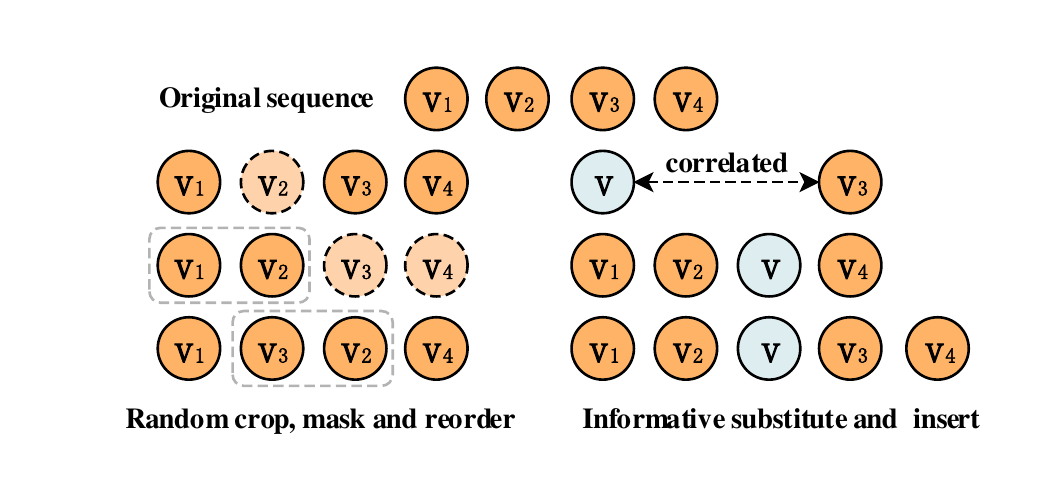}
	\vspace{-0.5cm}
	\caption{An example of the original sequence with different data augmentation operations.} 
	\label{fig3}
\end{figure}

\textbf{Data Augmentation Layer.} We first review and formulate existing random augmentation methods (i.e., random item crop, random item mask random item reorder) \cite{xie2020contrastive} and two informative augmentation methods (i.e, informative items substitute and informative items insert) \cite{liu2021contrastive}. 
We also run an example to explain how to augment sequences based on their length. This section assumes the original sequence is $s_u=[v_1, v_2,v_3,v_4]$. We illustrate the data augmentation Operators in Figure \ref{fig3}. 

\textbf{Learnable Model Augmentation Layer} 
In this section, we introduce a learnable model augmentation method, which can perform further data augmentation operations on the data-enhanced sequence, thereby generating high-quality enhanced views while ensuring the critical information in the sequence. For the selection of positive and negative samples, we take the enhanced view of the same sequence as the positive sample and the improved view of the different sequence as the negative sample. This work adopts learnable dropout SASRec as the sequence encoder. We use LBD in every FFN layer during training, which involves masking via ARM. As such, we generate a pair of views from model perspectives,  which can gain more comprehensive supervised signals.

\textbf{Embedding Layer}
For simplicity, the user embedding model chooses the learnable dropout SASRec method in our work. The sequence of user $u$ will be represented as $S^{m_i}_u$ and $S^{m_j}_u$ through the augmentation of the learnable model.

\textbf{Self-Supervised Loss Function}
Other than the next-item prediction, we can leverage other pretext tasks over the sequence to optimize the encoder, which harnesses self-supervised signals within the sequences. This paper investigates the widely adopted contrastive SSL scheme. This scheme constructs positive and negative view pairs from sequences and employs contrastive loss to optimize the encoder. We formulate the SSL step as follows:
\begin{equation}
\label{eq.8}
L_{ssl}(u^{m_i}_{u},u^{m_j} _{u})=  \\
-log\frac{exp(sim(u^{m_i}_{u},u^{m_j} _{u}))}{ {\textstyle \sum_{m=1}^{2N}}exp(sim(u^{m_i} _{u},u^{m_j} _{u})) },
\end{equation}
where $u^{m_i}_{u}$ and $u^{m_j} _{u}$ denote two views constructed for sequence $s_u$. The value of $m$ is not equal to $u^{m_i}_{u}$. $sim( )$ is the similarity function, like dot-product. Since each sequence has two views, we have $2N$ samples in a batch with $N$ sequences for training. The nominator indicates the agreement maximization between the positive pairs, while the denominator is interpreted as pushing away those negative pairs. 

\subsection{Joint Learning}
Because next-item prediction and SSL both model item relationships in sequences, to boost sequential recommendation performance with the SSL pretext objective function, we leverage the joint learning to optimize the loss function as follows:
\begin{equation}
\label{eq.9}
L_{total}=L_{rs}+\lambda L_{ssl},
\end{equation}
where $\lambda$ is a hyper-parameter that controls the weight of SSL.


\begin{table}[t]
\renewcommand{\arraystretch}{1.05}
\renewcommand\tabcolsep{1.3pt}
\centering
\caption{Datasets statistics.}
\label{datasets}
\begin{tabular}{c|ccccc}
\hline
Dataset   & users  & items  & interactions  & avg.length  & sparsity \\  \hline
Sports    & 35,598    & 18,357   & 296,337   & 8.3    & 99.95$\%$ \\ 
Toys      & 19,412    & 11,924   & 167,597   & 4.3    & 99.93$\%$ \\ 
Yelp      & 30,431    & 20,033   & 316,354   & 10.3   & 99.95$\%$  \\ 
\hline
\end{tabular}
\end{table}

\begin{table*}[t]
\renewcommand{\arraystretch}{1.0}
\renewcommand\tabcolsep{1.0pt}
\footnotesize
\caption{Performance comparisons of different methods. The best score is bold in each row, and the second-best is underlined.}
\label{Overall Comparison}
\centering
\begin{tabular}{p{1.25cm}<{\centering}|p{1.6cm}<{\centering}|p{1.25cm}<{\centering}p{1.25cm}<{\centering}|p{1.25cm}<{\centering}p{1.25cm}<{\centering}p{1.25cm}<{\centering}|p{1.25cm}<{\centering}p{1.25cm}<{\centering}p{1.25cm}<{\centering}|p{1.25cm}<{\centering}}
\hline
Dataset & Metric 
          & POP    & BPR    &GRU4Rec & Caser  & SASRec & S$^3$-Rec & CL4SRec & CoSeRec             & LMA4Rec   \\ \hline
\multirow{6}{*}{Sports} 
&HR@5     &0.0057  &0.0123  &0.0243  &0.0231  &0.0214  &0.0209  &0.0227  &\underline{0.0283}        &\textbf{0.0297}\\
&HR@10    &0.0091  &0.0215  &0.0339  &0.0303  &0.0345  &0.0322  &0.0358  &\underline{0.0429}        &\textbf{0.0439}\\
&HR@20    &0.0175  &0.0369  &0.0470  &0.0421  &0.0549  &0.0496  &0.0556  &\underline{0.0625}        &\textbf{0.0634}\\
&NDCG@5   &0.0041  &0.0076  &0.0113  &0.0129  &0.0143  &0.0137  &0.0143  &\underline{0.0192}        &\textbf{0.0199}\\
&NDCG@10  &0.0052  &0.0105  &0.0134  &0.0156  &0.0184  &0.0173  &0.0185  &\underline{0.0239}        &\textbf{0.0246}\\
&NDCG@20  &0.0073  &0.0144  &0.0156  &0.0203  &0.0236  &0.0217  &0.0234  &\underline{0.0288}        &\textbf{0.0293}\\
\hline
\multirow{6}{*}{Toys} 
&HR@5     &0.0065  &0.0122  &0.0396  &0.0403  &0.0378  &0.0461  &0.0435  &\textbf{0.0623}           &0.0619   \\
&HR@10    &0.0090  &0.0197  &0.0533  &0.0563  &0.0617  &0.0680  &0.0619  &\underline{0.0865}        &\textbf{0.0876}\\
&HR@20    &0.0143  &0.0327  &0.0699  &0.0768  &0.0906  &0.0964  &0.0829  &\underline{0.1162}        &\textbf{0.1188}\\
&NDCG@5   &0.0044  &0.0076  &0.0194  &0.0191  &0.0242  &0.0314  &0.0246  &\underline{0.0435}        &\textbf{0.0437}\\
&NDCG@10  &0.0052  &0.0100  &0.0224  &0.0359  &0.0319  &0.0385  &0.0305  &\underline{0.0513}        &\textbf{0.0521}\\
&NDCG@20  &0.0065  &0.0132  &0.0252  &0.0386  &0.0420  &0.0456  &0.0358  &\underline{0.0588}        &\textbf{0.0600}\\
\hline
\multirow{6}{*}{Yelp} 
&HR@5     &0.0056  &0.0185  &0.0171  &0.0115  &0.0190  &0.0162  &0.0204  &\underline{0.0225}        &\textbf{0.0233}\\
&HR@10    &0.0096  &0.0319  &0.0249  &0.0182  &0.0328  &0.0272  &0.0336  &\underline{0.0379}        &\textbf{0.0387}\\
&HR@20    &0.0158  &0.0525  &0.0363  &0.0291  &0.0543  &0.0463  &0.0583  &\underline{0.0626}        &\textbf{0.0636}\\
&NDCG@5   &0.0036  &0.0116  &0.0008  &0.0107  &0.0123  &0.0100  &0.0123  &\underline{0.0142}        &\textbf{0.0147}\\
&NDCG@10  &0.0049  &0.0159  &0.0096  &0.0153  &0.0167  &0.0136  &0.0174  &\underline{0.0191}        &\textbf{0.0196}\\
&NDCG@20  &0.0065  &0.0211  &0.0115  &0.0202  &0.0221  &0.0184  &0.0229  &\underline{0.0253}        &\textbf{0.0258}\\
\hline
\end{tabular}
\end{table*}

\section{Experiments}
In this section, We conduct plentiful experiments on datasets and answer the following research questions (\textbf{RQ}s): 

\textbf{RQ1}: How does LMA4Rec perform on sequential recommendation, comparing with existing baselines? 

\textbf{RQ2}: Are the key components in LMA4Rec necessary for improving the sequential recommendation performance? 

\textbf{RQ3}: Is LMA4Rec method sensitive to parameters? How do key hyper-parameters affect model performance? 

 \subsection{Dataset}
We conduct experiments on three public datasets, two Amazon review data in Sport and Toys categories respetively\footnote{http://jmcauley.ucsd.edu/data/amazon/links.html} and the Yelp dataset for the business recommendation \footnote{https://www.yelp.com/dataset}. We follow common practise in \cite{liu2021contrastive} to only keep the ‘5-core’ sequences. Table \ref{datasets} summarizes the detailed information of the public datasets.

\subsection{Baselines}
We include three groups of methods as baselines for comprehensive comparisons.

\textit{The first group baselines are non-sequential methods.} 
\textbf{POP} is based on the popularity of items. 
\textbf{BPR} \cite{rendle2012bpr} is a matrix factorization model with a pairwise Bayesian Personalized Ranking loss.
\textit{The second group baselines are sequential methods that use different deep neural architectures to encode sequences.} 
\textbf{GRU4Rec} \cite{hidasi2018recurrent} is a RNN-based method to model user sequences, which uses a new class of loss functions and sampling strategies to improve performance. 
\textbf{Caser} \cite{tang2018personalized} is a CNN-based method, which uses convolution to extract short-term preferences.
\textbf{SASRec} \cite{kang2018self} is a self-attention-based method to model user sequences, which capture users' dynamic interests via SA modules. 
\textit{The third group baselines additionally leverage the SSL objectives.}
\textbf{S$^3$-Rec} \cite{zhou2020s3} uses contrastive SSL with mask data augmentation to fuse correlation-ships among items. 
\textbf{CL4SRec}\cite{xie2020contrastive} maximizes agreements between two sequences augmentation. 
\textbf{CoSeRec} \cite{liu2021contrastive} improves the robustness of data augmentation under the SSL model by leveraging item-correlations.

\subsection{Evaluation and Implementation Details}
We refer to the paper \cite{liu2021contrastive} to evaluate models’ performance based on the whole items and report standard Hit Ratio $@K$ (HR$@K$) and Normalized Discounted Cumulative Gain $@K$ (NDCG$@K$) on all three datasets, where $K = \left\{5, 10, 20\right\}$. For commonly used parameters, we make the number of attention headers and self-attention blocks are 2, the maximum sequence length is 50 and the number of embedding size is 64. We adjust the parameters of the data augmentation part according to the range given in \cite{liu2021contrastive}. The model is optimized by the Adam optimizer. The value range of the hyperparameter $\lambda$ is $\left \{ 0,0.1,0.2,0.3,0.4,0.5 \right \} $, and the value range of hidden size is $\left \{ 64,128,192,256,320 \right \} $.

\begin{table*}[t]
\centering
\renewcommand{\arraystretch}{1.1}
\renewcommand\tabcolsep{1.0pt}
\footnotesize
\caption{Performance comparisons of different components.}
\begin{tabular}{p{3cm}<{\centering}
|p{1.0cm}<{\centering}p{1.0cm}<{\centering}p{1.0cm}<{\centering}p{1.0cm}<{\centering}
|p{1.0cm}<{\centering}p{1.0cm}<{\centering}p{1.0cm}<{\centering}p{1.0cm}<{\centering}
|p{1.0cm}<{\centering}p{1.0cm}<{\centering}p{1.0cm}<{\centering}p{1.0cm}<{\centering}}
\hline
\multirow{3}{*}{Metrics}     
& \multicolumn{4}{c|}{Sports}   & \multicolumn{4}{c|}{Toys}  & \multicolumn{4}{c}{Yelp} \\
\cline{2-13}        
        & \multicolumn{2}{c}{HR}     & \multicolumn{2}{c|}{NDCG} 
        & \multicolumn{2}{c}{HR}     & \multicolumn{2}{c|}{NDCG}  
        & \multicolumn{2}{c}{HR}     & \multicolumn{2}{c}{NDCG} \\
\cline{2-13}
        & @5     & @10     & @5     & @10       
        & @5     & @10     & @5     & @10     
        & @5     & @10     & @5     & @10       \\
\hline
LMA4Rec & \textbf{0.0297}   & \textbf{0.0439}   & \textbf{0.0199}   & \textbf{0.0246}          & \textbf{0.0619}   & \textbf{0.0876}   & \textbf{0.0437}   & \textbf{0.0521}
        & \textbf{0.0233}   & \textbf{0.0387}   & \textbf{0.0147}   & \textbf{0.0196}\\
        
LMA4Rec $w/o$ SSL    & 0.0227    & 0.0342    & 0.0155     & 0.0192     
                     & 0.0513    & 0.0726    & 0.0339     & 0.0408     
                     & 0.0174    & 0.0288    & 0.0108     & 0.0145 \\
                     
LMA4Rec $w/o$ LMA    & 0.0283    & 0.0429    & 0.0192     & 0.0239                                          & 0.0623    & 0.0865    & 0.0435     & 0.0513
                     & 0.0225    & 0.0379    & 0.0142     & 0.0191 \\ 
                     
LMA4Rec $w/o$ DA     & 0.0140    & 0.0236    & 0.0090     & 0.0120         
                     & 0.0509    & 0.0703    & 0.0348     & 0.0410
                     & 0.0195    & 0.0305    & 0.0132     & 0.0167 \\ 
\hline
\end{tabular}
\label{Ablation Study}
\end{table*}

\subsection{Overall Comparisons (\textbf{RQ1})}
To verify the effectiveness of our proposed LMA4Rec method in the sequence recommendation task, to ensure the fairness of the models, we use the same data partition rules for all models and then apply them to the sequence recommendation task. Our purpose is to predict the possible item of the user at the time of $t$ based on the interaction sequence before the time of the user's visit at the time of $t-1$. Table \ref{Overall Comparison} shows the experimental results of the LMA4Rec method and its baselines. After comparing and analyzing the results, we can draw the following conclusions:

(1) The performance of the non-sequential models is worse than that of the sequential recommendation methods. The results show the importance of sequential mining patterns for the next prediction. Then,  for the sequence model, the self-attention-based method achieves better performance than the recommendation model of CNN-based and RNN-based baselines, which shows that the self-attention networks can more effectively capture the sequence pattern.

(2)  As for self-supervised methods, S$^3$-Rec uses self-supervised signals by masking items in the order, but its performance is worse than SASRec. There are two reasons to explain this performance: On the one hand, S$^3$-Rec has independent two-stage training; on the other hand, there is no information augmentation part in S$^3$-Rec, and weak supervision signals are obtained. CL4SRec always performs better than non-sequential and sequential baselines. The results verify using data augmentation and contrast SSL in sequential recommendations. However, CL4SRec still performs worse than CoSeRec. This is because CL4SRec has no information augmentation. Our LMA4Rec achieves better recommendation results than CL4SRec and CoSeRec. This is because LMA4Rec uses model augmentation to improve recommendation performance.

(3) Our proposed LMA4Rec is superior to other models on all three datasets in terms of most evaluation indicators. Compared with the best baseline CoseRec, LMA4Rec achieves improvements in terms of HR and NDCG. Our LMA4Rec method has achieved varying degrees of improvement. For example, on the Sports dataset, its improvements in terms of HR$@5,10,20$ are: 4.95$\%$, 2.33$\%$, 1.44$\%$ and those in terms of NDCG$@5,10,20$ are: 3.65$\% $, 2.93$\%$, and 1.74$\%$.
Because in actual applications, we have observed very little user interaction data, SSL based on the learnable model generates two views of the same user's sequence. Adding increments in the training process can increase the number of training, thereby alleviating the sparseness of the data; on the other hand, the contrast loss improves the robustness of the model.
\subsection{Influence of Components (\textbf{RQ2})}
In this section, We implement ablation studies to evaluate the effectiveness of each part of LMA4Rec. Specifically, we perform four ablation experiments as follows:

\textbf{(1)} LMA4Rec: Nothing removed, using the learnable model augmentation module, data augmentation and the self-supervised learning module for SR. 
\textbf{(2)} LMA4Rec $w/o$ SSL: Remove the self-supervised learning module, using the learnable dropout SASRec for SR.
\textbf{(3)} LMA4Rec $w/o$ LMA: Remove the learnable model augmentation module, using the self-supervised learning module and data augmentation module for SR.
\textbf{(4)} LMA4Rec $w/o$ DA: Remove the data augmentation module, using the self-supervised learning module and learnable model augmentation module for SR.

Table \ref{Ablation Study} lists the results of the ablation study, showing how each part affects the final performance of LMA4Rec. 
For the Amazon dataset and Yelp dataset, the best results are achieved when all components are present. When different parts are removed, the recommendation effect decreases to varying degrees. 
Obviously, the self-supervised learning module, the learnable model augmentation module and the data augmentation module can all help improve the performance of LAM-SSL. 
The experimental results proved that high-quality positive samples could be generated via learnable model augmentation, which is conducive to learning SSL tasks, thereby improving the quality and accuracy of recommendations.

\subsection{Influence of Hyper-parameters (\textbf{RQ3})}
\begin{figure}[t]
	\centering
	\includegraphics[width=0.73\textwidth]{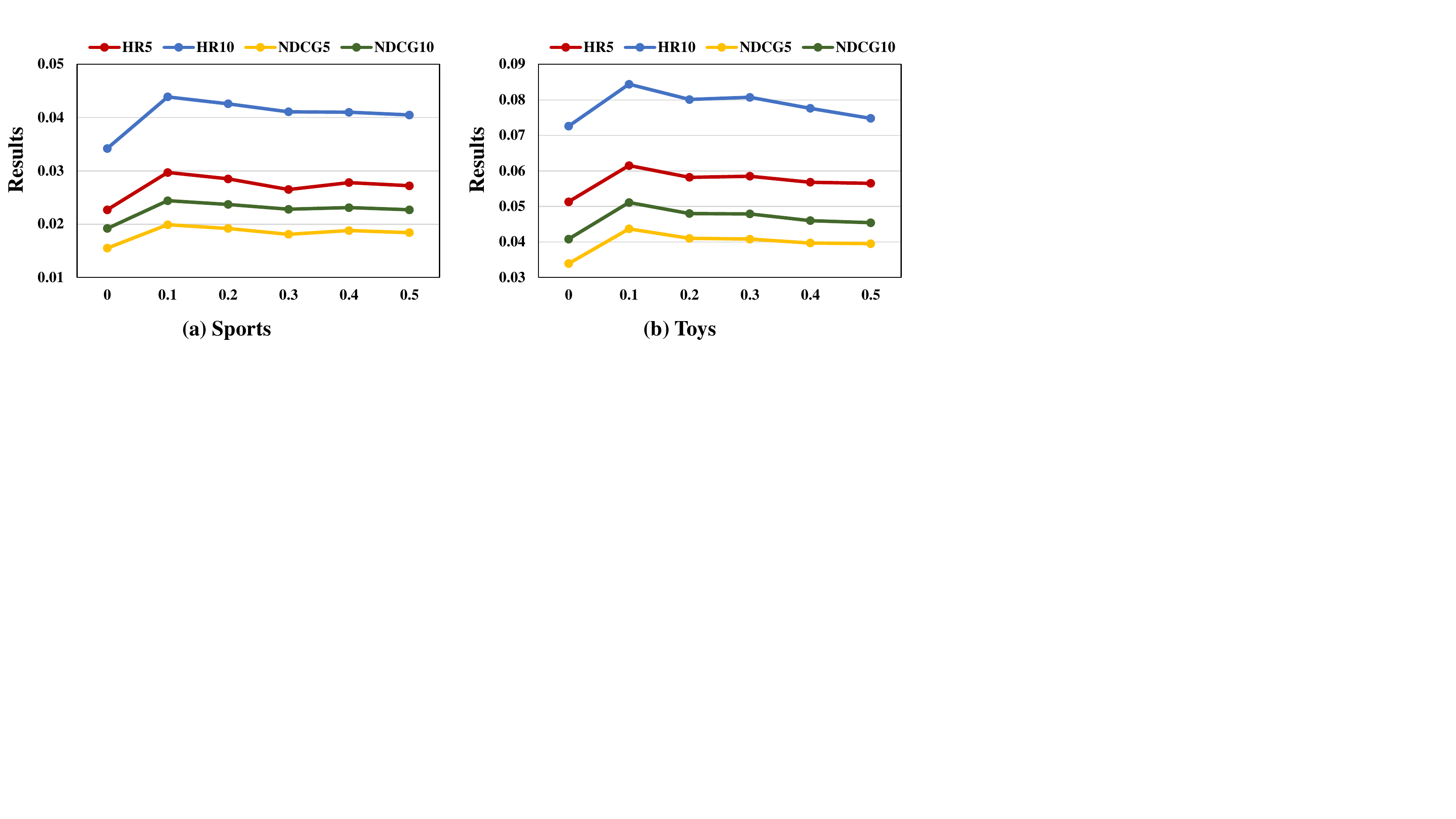}
	\vspace{-4.9cm}
	\caption{The influence of $\lambda$.} \label{fig4}
\end{figure}

\begin{figure}[t]
	\centering
	\includegraphics[width=0.73\textwidth]{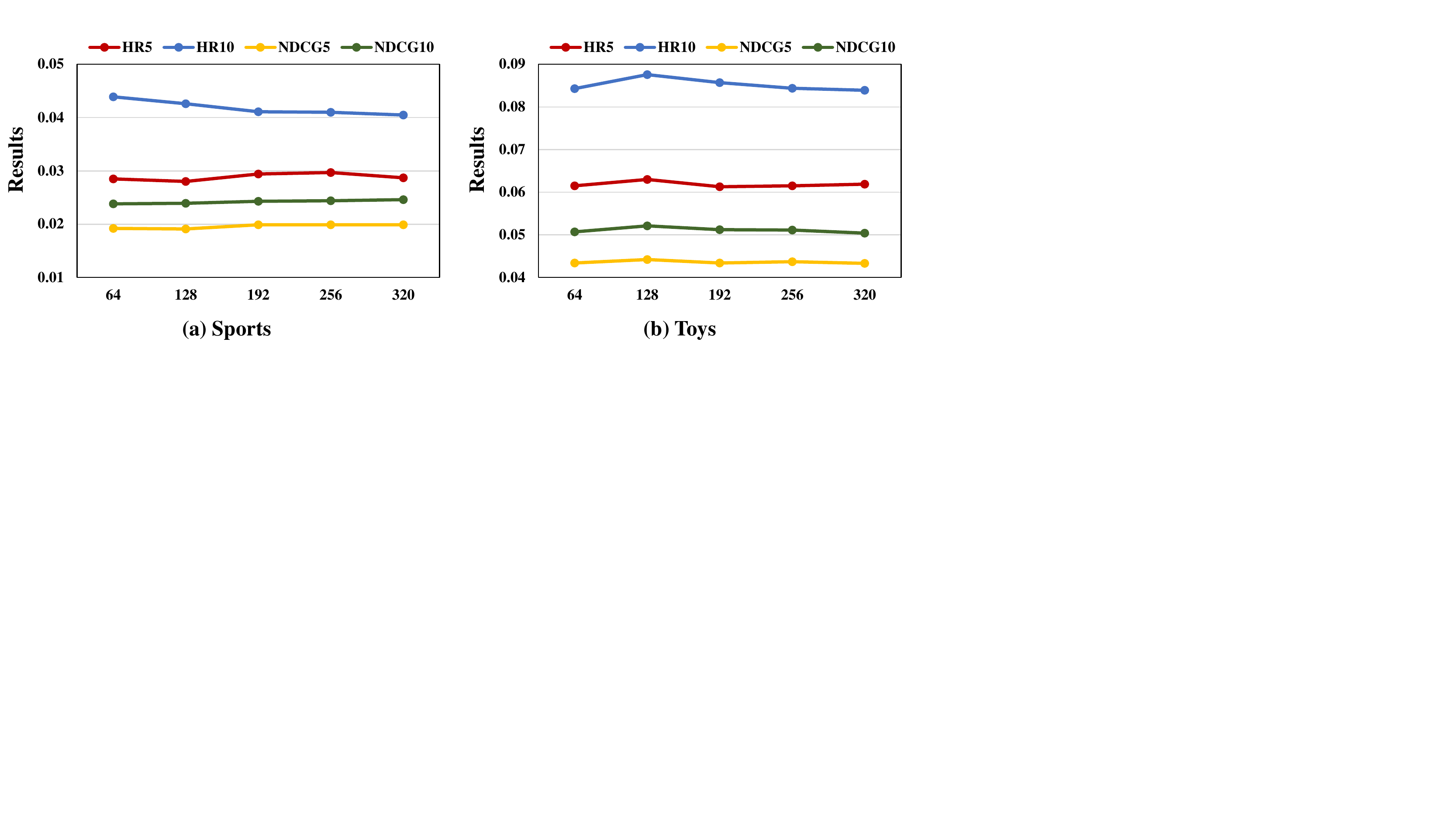}
	\vspace{-4.9cm}
	\caption{The influence of the hidden size.} \label{fig5}
\end{figure}

The essential hyper-parameters in LMA4Rec are $\lambda$ and hidden size. Specifically, we fix one parameter and adjust the value of another hyper-parameter. We show the sequential recommendation results on two different datasets and draw line charts to show their impacts as shown in Figure \ref{fig4} and \ref{fig5}.

\textbf{The influence of the $\lambda$:} We adjusted the lambda from 0 to 0.5 with a step size of 0.1. Experimental results are shown in Figure \ref{fig4}. From Figure \ref{fig4}, we can see that the recommendation performance changes with the change of the lambda value. LMA4Rec achieves the best results effect when the lambda is 0.1 on both the Sports and the Toys datasets. When the proportion of self-supervision and sequential recommendation is closer, the accuracy of recommendation will be reduced.

\textbf{The influence of the hidden size:} We adjusted the hidden size from 64 to 320 and used 64 as the step size. Experimental results are shown in Figure \ref{fig5}. From Figure \ref{fig5}, it can be observed that best the hidden size is different in different datasets. As for the Sports dataset, the best results are obtained with the hidden size is 256. On the Toys dataset, the best results are gained with the hidden size is 128. Moreover, the effects on the two datasets do not fluctuate greatly, so we can speculate that the hidden size has a limited impact on the recommendation performance.

\section{Conclusion}
This work proposed a novel self-supervised learning paradigm, which simultaneously employs model and data augmentation to constructive views. We adopted this paradigm to sequential recommendation and proposed a new model called LMA4Rec. We conducted comprehensive experiments and our experimental results show that the LMA4Rec method effectively improves sequential recommendation performance compared with baseline methods. Moreover, ablation studies with respect to learnable model augmentation methods also demonstrate the validity of every component in LMA4Rec.

\bibliographystyle{named}
\bibliography{ijcai22}

\begin{thebibliography}{}

\bibitem[\protect\citeauthoryear{Boluki \bgroup \em et al.\egroup
  }{2020}]{boluki2020learnable}
Shahin Boluki, Randy Ardywibowo, Siamak~Zamani Dadaneh, Mingyuan Zhou, and
  Xiaoning Qian.
\newblock Learnable bernoulli dropout for bayesian deep learning.
\newblock In {\em AISTATS}, pages 3905--3916, 2020.

\bibitem[\protect\citeauthoryear{Chen \bgroup \em et al.\egroup
  }{2020}]{chen2020simple}
Ting Chen, Simon Kornblith, Mohammad Norouzi, and Geoffrey Hinton.
\newblock A simple framework for contrastive learning of visual
  representations.
\newblock In {\em ICML}, pages 1597--1607, 2020.

\bibitem[\protect\citeauthoryear{Devlin \bgroup \em et al.\egroup
  }{2019}]{devlin2018bert}
Jacob Devlin, Ming{-}Wei Chang, Kenton Lee, and Kristina Toutanova.
\newblock {BERT:} pre-training of deep bidirectional transformers for language
  understanding.
\newblock In {\em NAACL-HLT}, pages 4171--4186, 2019.

\bibitem[\protect\citeauthoryear{Gao \bgroup \em et al.\egroup
  }{2021}]{gao2021simcse}
Tianyu Gao, Xingcheng Yao, and Danqi Chen.
\newblock Simcse: Simple contrastive learning of sentence embeddings.
\newblock In {\em EMNLP}, pages 6894--6910, 2021.

\bibitem[\protect\citeauthoryear{He and McAuley}{2016a}]{he2016fusing}
Ruining He and Julian McAuley.
\newblock Fusing similarity models with markov chains for sparse sequential
  recommendation.
\newblock In {\em ICDM}, pages 191--200, 2016.

\bibitem[\protect\citeauthoryear{He and McAuley}{2016b}]{Fossil}
Ruining He and Julian~J. McAuley.
\newblock Fusing similarity models with markov chains for sparse sequential
  recommendation.
\newblock In {\em ICDM}, pages 191--200, 2016.

\bibitem[\protect\citeauthoryear{He \bgroup \em et al.\egroup
  }{2020}]{he2020momentum}
Kaiming He, Haoqi Fan, Yuxin Wu, Saining Xie, and Ross Girshick.
\newblock Momentum contrast for unsupervised visual representation learning.
\newblock In {\em CVPR}, pages 9729--9738, 2020.

\bibitem[\protect\citeauthoryear{Hidasi and
  Karatzoglou}{2018}]{hidasi2018recurrent}
Bal{\'a}zs Hidasi and Alexandros Karatzoglou.
\newblock Recurrent neural networks with top-k gains for session-based
  recommendations.
\newblock In {\em CIKM}, pages 843--852, 2018.

\bibitem[\protect\citeauthoryear{Hidasi \bgroup \em et al.\egroup
  }{2016}]{hidasi2015session}
Bal{\'a}zs Hidasi, Alexandros Karatzoglou, Linas Baltrunas, and Domonkos Tikk.
\newblock Session-based recommendations with recurrent neural networks.
\newblock {\em ICLR}, 2016.

\bibitem[\protect\citeauthoryear{Kang and McAuley}{2018}]{kang2018self}
Wang-Cheng Kang and Julian McAuley.
\newblock Self-attentive sequential recommendation.
\newblock In {\em ICDM}, pages 197--206, 2018.

\bibitem[\protect\citeauthoryear{Liu \bgroup \em et al.\egroup
  }{2021a}]{liu2021social}
Yuejiang Liu, Qi~Yan, and Alexandre Alahi.
\newblock Social nce: Contrastive learning of socially-aware motion
  representations.
\newblock In {\em ICCV}, pages 15118--15129, 2021.

\bibitem[\protect\citeauthoryear{Liu \bgroup \em et al.\egroup
  }{2021b}]{liu2021contrastive}
Zhiwei Liu, Yongjun Chen, Jia Li, Philip~S. Yu, Julian~J. McAuley, and Caiming
  Xiong.
\newblock Contrastive self-supervised sequential recommendation with robust
  augmentation.
\newblock {\em CoRR}, abs/2108.06479, 2021.

\bibitem[\protect\citeauthoryear{Rendle \bgroup \em et al.\egroup
  }{2009}]{rendle2012bpr}
Steffen Rendle, Christoph Freudenthaler, Zeno Gantner, and Lars Schmidt-Thieme.
\newblock Bpr: Bayesian personalized ranking from implicit feedback.
\newblock {\em UAI}, pages 452--461, 2009.

\bibitem[\protect\citeauthoryear{Rendle \bgroup \em et al.\egroup
  }{2010}]{rendle2010factorizing}
Steffen Rendle, Christoph Freudenthaler, and Lars Schmidt-Thieme.
\newblock Factorizing personalized markov chains for next-basket
  recommendation.
\newblock In {\em WWW}, pages 811--820, 2010.

\bibitem[\protect\citeauthoryear{Tang and Wang}{2018}]{tang2018personalized}
Jiaxi Tang and Ke~Wang.
\newblock Personalized top-n sequential recommendation via convolutional
  sequence embedding.
\newblock In {\em WSDM}, pages 565--573, 2018.

\bibitem[\protect\citeauthoryear{van~den Oord \bgroup \em et al.\egroup
  }{2016}]{oord2016conditional}
A{\"{a}}ron van~den Oord, Nal Kalchbrenner, Lasse Espeholt, Koray Kavukcuoglu,
  Oriol Vinyals, and Alex Graves.
\newblock Conditional image generation with pixelcnn decoders.
\newblock In {\em NIPS}, pages 4790--4798, 2016.

\bibitem[\protect\citeauthoryear{Wu \bgroup \em et al.\egroup
  }{2017}]{wu2017recurrent}
Chao-Yuan Wu, Amr Ahmed, Alex Beutel, Alexander~J Smola, and How Jing.
\newblock Recurrent recommender networks.
\newblock In {\em WSDM}, pages 495--503, 2017.

\bibitem[\protect\citeauthoryear{Wu \bgroup \em et al.\egroup
  }{2021}]{wu2021self}
Jiancan Wu, Xiang Wang, Fuli Feng, Xiangnan He, Liang Chen, Jianxun Lian, and
  Xing Xie.
\newblock Self-supervised graph learning for recommendation.
\newblock In {\em SIGIR}, pages 726--735, 2021.

\bibitem[\protect\citeauthoryear{Xie \bgroup \em et al.\egroup
  }{2020}]{xie2020contrastive}
Xu~Xie, Fei Sun, Zhaoyang Liu, Jinyang Gao, Bolin Ding, and Bin Cui.
\newblock Contrastive pre-training for sequential recommendation.
\newblock {\em CoRR}, abs/2010.14395, 2020.

\bibitem[\protect\citeauthoryear{Xu \bgroup \em et al.\egroup
  }{2019}]{xu2019graph}
Chengfeng Xu, Pengpeng Zhao, Yanchi Liu, Victor~S Sheng, Jiajie Xu, Fuzhen
  Zhuang, Junhua Fang, and Xiaofang Zhou.
\newblock Graph contextualized self-attention network for session-based
  recommendation.
\newblock In {\em IJCAI}, volume~19, pages 3940--3946, 2019.

\bibitem[\protect\citeauthoryear{Yang \bgroup \em et al.\egroup
  }{2019}]{yang2019xlnet}
Zhilin Yang, Zihang Dai, Yiming Yang, Jaime Carbonell, Russ~R Salakhutdinov,
  and Quoc~V Le.
\newblock Xlnet: Generalized autoregressive pretraining for language
  understanding.
\newblock {\em NIPS}, 32, 2019.

\bibitem[\protect\citeauthoryear{Yin and Zhou}{2019}]{yin2018arm}
Mingzhang Yin and Mingyuan Zhou.
\newblock {ARM:} augment-reinforce-merge gradient for stochastic binary
  networks.
\newblock In {\em ICLR}, 2019.

\bibitem[\protect\citeauthoryear{You \bgroup \em et al.\egroup
  }{2020}]{you2020graph}
Yuning You, Tianlong Chen, Yongduo Sui, Ting Chen, Zhangyang Wang, and Yang
  Shen.
\newblock Graph contrastive learning with augmentations.
\newblock {\em NIPS}, 33:5812--5823, 2020.

\bibitem[\protect\citeauthoryear{Zhao \bgroup \em et al.\egroup
  }{2019}]{zhao2020go}
Pengpeng Zhao, Haifeng Zhu, Yanchi Liu, Jiajie Xu, Zhixu Li, Fuzhen Zhuang,
  Victor~S. Sheng, and Xiaofang Zhou.
\newblock Where to go next: {A} spatio-temporal gated network for next {POI}
  recommendation.
\newblock In {\em AAAI}, pages 5877--5884, 2019.

\bibitem[\protect\citeauthoryear{Zhou \bgroup \em et al.\egroup
  }{2020}]{zhou2020s3}
Kun Zhou, Hui Wang, Wayne~Xin Zhao, Yutao Zhu, Sirui Wang, Fuzheng Zhang,
  Zhongyuan Wang, and Ji-Rong Wen.
\newblock S3-rec: Self-supervised learning for sequential recommendation with
  mutual information maximization.
\newblock In {\em CIKM}, pages 1893--1902, 2020.

\bibitem[\protect\citeauthoryear{Zhu \bgroup \em et al.\egroup
  }{2017}]{zhu2017next}
Yu~Zhu, Hao Li, Yikang Liao, Beidou Wang, Ziyu Guan, Haifeng Liu, and Deng Cai.
\newblock What to do next: Modeling user behaviors by time-lstm.
\newblock In {\em IJCAI}, pages 3602--3608, 2017.

\end{thebibliography}

\end{document}